\RequirePackage{fix-cm}
\documentclass[twocolumn,epjc3]{svjour3}  
\smartqed  
\RequirePackage{graphicx}
\journalname{Eur. Phys. J. B}
\begin{document}

\title{Multifractality and the distribution of the Kondo temperature at the Anderson transition
}
\subtitle{}


\author{Keith Slevin\thanksref{e1,addr1}
        \and
        Stefan Kettemann\thanksref{addr2} 
        \and
        Tomi Ohtsuki\thanksref{addr4}
}

\thankstext{e1}{e-mail: slevin@phys.sci.osaka-u.ac.jp}


\institute{Department of Physics, Graduate School of Science, Osaka University, Toyonaka, Osaka 560-0043, Japan\label{addr1}
           \and
           Department of Physics and Earth Sciences, School of Engineering and Science, Jacobs University Bremen, Bremen 28759, Germany\label{addr2}
           \and
        Physics Division, Sophia University, Chiyoda, Tokyo 102-8554, Japan\label{addr4}
}

\date{Received: date / Accepted: date}

\maketitle

\begin{abstract}
Using numerical simulations, we investigate the distribution of Kondo temperatures at the Anderson transition. 
In agreement with previous work, we find that the distribution has a long tail at small Kondo temperatures. 
Recently,  an approximation for the tail of the distribution was derived analytically. 
This approximation takes into account the multifractal distribution of the wavefunction amplitudes (in the parabolic approximation),
and power law correlations between wave function intensities, 
at the Anderson transition. 
It was predicted that the  distribution of Kondo temperatures has a power law tail with a universal exponent. 
Here, we attempt to check that this prediction holds in a numerical simulation of Anderson's model of localisation in three dimensions.
\end{abstract}

\section{Introduction}
\label{intro}

At low temperatures the magnetic moment of a magnetic 
impurity in a metal is screened by the exchange interaction with the conduction electrons\cite{Kondo64,Wilson75}.
Above the Kondo temperature $T_{\rm K}$ the magnetic impurity contributes a Curie like
term to the magnetic susceptibility.
Below $T_{\rm K}$ the magnetic moment is screened and the contribution to the susceptibility 
is a temperature independent Pauli like contribution.

Following Nagaoka\cite{Nagaoka65}  and Suhl\cite{Suhl65}, for the simplest model of a 
non-disordered (i.e. clean) metal, with a band of width $D$ and a
position and energy independent local density of states (LDOS)
\begin{equation}
    \rho = \frac{1}{D}\,,
\end{equation}
the Kondo temperature $T_{\rm K}$ is approximately
\begin{equation}
    T_{\rm K} \approx 0.567 D \exp \left( - \frac{1}{\rho J}\right)\,.
\end{equation}
Here, $J$ is the constant describing the exchange coupling $-J\vec{S}\cdot \vec{s}$ 
of the spin $\vec{S}$ of the
magnetic impurity with the spin $\vec{s}$ of a conduction electron.

In a disordered metal the LDOS exhibits strong fluctuations as a function of
both position and energy. These fluctuations are reflected in fluctuations
of the Kondo temperature.
Previous work has shown that the Kondo temperature has a very broad
distribution and, in particular, that there is a long tail at low Kondo temperatures\cite{Dobrosavljevic92,Cornaglia06,Kettemann06,Kettemann12,Lee14}.
The fluctuations in the LDOS reflect the spatial fluctuations in the
eigenstates $\psi\left(\vec{r}\right)$ of the conduction electrons.
These spatial fluctuations of the eigenstates are also reflected in   
various other phenomena such as the broad distribution of
conductance\cite{Shapiro87,Slevin01} and the enhancement of the superconducting critical temperature\cite{Feigelman07,Burmistrov12}.

At the Anderson transition the fluctuations of the eigenfunction
intensities are multifractal and described by a multifractal spectrum $f\left(\alpha\right)$\cite{Evers08,Rodriguez08,Vasquez08,Rodriguez09}.
In terms of the multifractal spectrum, the probability distribution 
\begin{equation}
    P\left( \alpha \right) = p\left( \alpha \right) d\alpha,
\end{equation}
of the quantity
\begin{equation}
    \alpha = -\frac{\ln\left| \psi \right|^2}{\ln L}\,,
\end{equation}
is given by
\begin{equation}
    p\left( \alpha \right) \propto L ^{ f\left(\alpha\right) -d} \,.
\end{equation}
Here, $L$ is the linear size of the system and
\begin{equation}
    d=3\,,
\end{equation}
its dimensionality.
A rough approximation for the multifractal spectrum is the following parabolic 
form
\begin{equation} \label{f_parabolic}
 f\left(\alpha\right) \approx d - \frac{\left(\alpha - \alpha_0\right)^2}
 {4 \left( \alpha_0 - d \right)}\,.
\end{equation}
When required we use\cite{Rodriguez11}   
\begin{equation}\label{alpha0_value}
    \alpha_0 \approx 4.043\,,
\end{equation}
as a numerical estimate of $\alpha_0$ for the Anderson transition.
This value is in agreement with the later estimate of Ref. \cite{Ujfalusi15}.

The LDOS also involves correlations in the intensities of
different eigenstates.
Following Ref. \cite{Cuevas07},
when one of the states has energy equal to the energy of the mobility edge, 
the pairwise correlator 
\begin{equation} \label{correlations}
    C_{mn} = L^{d} \int d^dr\, \left\langle |\psi_m\left(\vec{r}\right)|^2
   |\psi_n\left(\vec{r}\right)|^2 \right\rangle\,,
\end{equation}
(where angular brackets indicates a disorder average) has the form
\begin{equation} 
C_{mn} = \left( \frac{E_{\rm c}}{ \max \left( \left| E_m-E_n\right|, \Delta \right)} \right) ^ {\eta / d}\,,
\end{equation}
when the energy difference does not exceed the correlation energy $E_{\rm c}$,
\begin{equation}
    \left| E_m-E_n \right| < E_{\rm c} \,,
\end{equation}
and
\begin{equation}
C_{mn} = \left( \frac{E_{\rm c}}{  \left| E_m-E_n\right| } \right) ^ {2} \,,
\end{equation}
when 
\begin{equation}
    \left| E_m-E_n \right| > E_{\rm c}\,.
\end{equation}
Here
\begin{equation}
    \Delta = \frac{D}{L^3} \,,
\end{equation}
is the approximate energy level spacing of the system.
For energy differences smaller than $E_{\rm c}$ the
correlations are enhanced compared with the plane wave limit, for which
$C_{mn}=1$.
Considering the limit 
\begin{equation}
    \left| E_m-E_n \right| \rightarrow \Delta \,,
\end{equation}
one recovers
\begin{equation}
    C_{mn} \rightarrow L^{2d} \left\langle |\psi_m\left(\vec{r}\right)|^4 \right\rangle\,.
\end{equation}
The scaling with system size of the right hand side of this equation 
is related to the fractal dimension $D_2$ (see Sec II C of Ref \cite{Evers08}),
which leads to
\begin{equation}
    \eta = d - D_2 \;.
\end{equation}
Using the parabolic approximation for the multifractal spectrum to 
calculate $D_2$ then gives
\begin{equation} \label{eta_parabolic_relation}
    \eta = 2\left(\alpha_0-d\right)\,.
\end{equation}
With the numerical estimate Eq. (\ref{alpha0_value}) for $\alpha_0$ we obtain
\begin{equation} \label{eta_parabolic_value}
    \eta \approx 2.086\,.
\end{equation}

\section{The distribution of Kondo temperatures at the Anderson transition}

In Ref. \cite{Kettemann12}, an approximation for the distribution of Kondo temperatures 
at the Anderson transition was derived incorporating both multifractality in the parabolic approximation Eq. (\ref{f_parabolic})  
and  pairwise power law correlations Eq. (\ref{correlations}).
For Kondo temperatures in the range 
\begin{equation}\label{temperature_range}
    \Delta < T_{K} \ll T_{\rm K}^{(0)} \,,
\end{equation}
where $T_{\rm K}^{(0)}$ is the Kondo temperature
of the clean system,
the result reads
\begin{equation}
    P(x) = p(x) dx\,,
\end{equation}
where $x$ is the variable
\begin{equation}\label{x_def}
    x = \frac{T_{\rm K}}{T^{(0)}_K}\,,
\end{equation}
and
\begin{eqnarray}\label{x_pdf}
    p(x) & \approx & A x^{(\eta/2d)-1} \\ \nonumber
    & & \times \exp \left\{- \frac{1}{2 c_{1}} 
    \left( \frac{T_{K}^{(0)}}{E_{\rm c}}\right)^{\frac{\eta}{d}}
    x^{\frac{\eta}{d}}
    \ln^2 \left[ x \right]
    \right\}.
\end{eqnarray}
In this formula $A$ is a normalisation constant, which we determine numerically.
The ratio of the Kondo temperature of the clean system $T_{\rm K}^{(0)}$ to the
correlation energy $E_{\rm c}$ also appears.
When evaluating this ratio we use the approximations\cite{Cuevas07}
\begin{equation}
    E_{\rm c}=\frac{D}{2\ln\left(2d\right)},
\end{equation}
and\cite{Kettemann12}
\begin{equation}
    T_{\rm K}^{(0)} = E_{\rm c} \exp \left( \frac{1}{2} - \frac{D}{J}\right)\,.
\end{equation}
The constant $c_1$ is given by the following definite integral
\begin{equation}
    c_1= \int_0^{\infty} \frac{du }{u} \int_0^{\infty} \frac{dv}{v}
\tanh(u/2) \tanh(u/2)   h\left(u,v\right)\,,
\end{equation}
where
\begin{equation}
    h\left(x,y\right) = |u - v|^{-(\eta/d)}+|u+v|^{-(\eta/d)}\,.
\end{equation}
A numerical integration yielded the estimate
\begin{equation}
    c_1 \approx 14.728\,.
\end{equation}

The distribution Eq.(\ref{x_pdf}) applies only to non-zero Kondo temperatures in the range given in Eq. (\ref{temperature_range}).
In a finite system, a certain fraction $n_{\rm FM}$ of the magnetic impurities are not screened even at zero temperature, i.e. they remain free magnetic moments. 
In clean systems such free moments only exist at exchange couplings smaller than
\begin{equation}
    J_{-} = D/(\ln (2 L^d) +C)\;,
\end{equation}
where $C=0.577...$ is Euler's constant.
However, in disordered systems,
due to fluctuations of the energy level spacing and local wave function intensities,
there are free moments even for $J>J_{-}$. 
In Ref. \cite{Kettemann12} it was found that, at the Anderson transition,
the dominant mechanism for the creation of magnetic moments for $J>J_{-}$ 
is the formation of 
local pseudo-gaps due to multifractal power law correlations. 
It was found that $n_{\rm FM}$ is proportional to the fraction of sites with $\alpha > \alpha_c$, where 
\begin{equation}
    \alpha_c = \alpha_0 + d \frac{J}{D} \;,
\end{equation}
is the critical value of $\alpha$ for which the Nagaoka-Suhl equation
(see Eqs. (\ref{NagaokaSuhl}) and  (\ref{Ftk}) in Sec. \ref{model_method} below)
has no solution\cite{Kettemann12,Kettemann09}.  
Using the  parabolic approximation for the  multifractal distribution then yields
\begin{equation}  \label{nfm}
n_{\rm FM}  = \frac{1-{\rm erf}((dJ/D)\sqrt{\ln L/(2 \eta)} )}{1-{\rm erf}(-\alpha_0\sqrt{\ln L/(2 \eta)})} \;.
\end{equation}
Here, ${\rm erf}\left(x\right)$ is the error function.
Note that we expect that the concentration of free magnetic moments 
scales to zero in the limit of infinite system size, i.e. that
\begin{equation}
   \lim_{L \rightarrow \infty}  n_{\rm FM} \rightarrow 0\;,
\end{equation}
at the Anderson transition.

For $\eta > 0$ and sufficiently large, and $J$ sufficiently small,
the low temperature tail of the Kondo temperature distribution
Eq. (\ref{x_pdf}) can be approximated as a power law 
\begin{equation}\label{x_powerlaw}
    p(x) \approx A x^{(\eta/2d)-1}\;,
\end{equation}
with a universal exponent
\begin{equation}\label{x_exponent_prediction}
    \frac{\eta}{2d}-1\;.
\end{equation}
In what follows, we attempt to verify in a numerical simulation of Anderson's model
of localisation in three dimensions that
the distribution of Kondo temperatures does indeed follow such a power law
and to verify the prediction for the exponent, 
Eq. (\ref{x_exponent_prediction}).

It is important to note that for Kondo  temperatures below the level spacing 
\begin{equation}
    T_{\rm K} < \Delta \;,
\end{equation}
the distribution is also power law but with a non-universal exponent 
that depends on the exchange coupling
\begin{equation}
     p(x) \propto x^{(J/D)-1} L^{-d^2 J^2/(2 \eta D^2)}\,.
\end{equation}
This should be borne in mind when looking at previous numerical work \cite{Cornaglia06,Kettemann06} where
system sizes were more limited.
Here, we simulate much larger system sizes and focus on the universal regime only.

\section{Model and method}\label{model_method}
We use Anderson's model of localisation \cite{Anderson58} to model the disordered system.
The Hamiltonian is
\begin{equation}\label{Hamiltonian}
H = \sum_{i}  \left| i \right> \epsilon_i\left< i \right| - \sum_{<ij>} \left| i \right> \left< j \right| \,.
\end{equation}
The ket $\left| i \right>$ represents an orbital localised on lattice site $i$.
The first sum is over all the sites of a three dimensional cubic lattice and the second sum is over nearest neighbour sites.
We impose periodic boundary conditions in all directions so that all lattice sites are statistically equivalent.
The unit of energy is fixed by taking the hopping energy between nearest neighbour orbitals as unity.
The orbital energies $\epsilon_i$ are independent and identically distributed random variables with a uniform distribution centred at zero and of width $W$. We
refer to the parameter $W$ as the disorder.
The Anderson transition occurs at a critical disorder
\begin{equation}
    W=W_{\rm c} \left (E_{\rm F} \right)\;,
\end{equation}
which is a function of the Fermi energy $E_{\rm F}$.
We work at the band centre
\begin{equation}
    E_{\rm F} = 0\;,
\end{equation}
and use the estimate of the critical disorder 
\begin{equation}
    W_{\rm c} \left (E_{\rm F} = 0 \right) \approx 16.54\;,
\end{equation}
given in Refs. \cite{Slevin14,Slevin18}.

We suppose that there is a single spin one-half magnetic impurity at some arbitrary site. This interacts with the conduction electrons through 
an on-site exchange coupling of magnitude $J$.
We approximate the Kondo temperature $T_{\rm K}$ by solving the one-loop equation of Nagaoka and Suhl\cite{Nagaoka65,Suhl65},
\begin{equation}\label{NagaokaSuhl}
     F\left( {{T_{\rm K}}} \right) = 1\;,
\end{equation}
where
\begin{equation}\label{Ftk}
    F\left( {{T_{\rm K}}} \right) = \frac{J}{2}\int {\rho \left( {E,\vec r} \right)\frac{{\tanh \left( {\left( {E - {E_F}} \right)/2{T_{\rm K}}} \right)}}{{E - {E_F}}}dE} \;.
\end{equation}
Here, $\rho(E,\vec{r})$ is the LDOS for energy $E$ at the lattice site $\vec{r}$ where the magnetic impurity is situated, and $E_F$ is the Fermi energy.
The LDOS is given by
\begin{equation}\label{LDOS}
\rho\left(E , \vec{r} \right) = \sum_{n} \left| \psi_n \left( \vec{r} \right)\right|^2 \delta \left( E - E_n \right)\,.
\end{equation}
The sum is over all eigenstates of the Hamiltonian.
Evaluation of this formula would require a full diagonalization of the Hamiltonian, which is  impractical.
Moreover, most of the information obtained in the diagonalization would not be required, since while we need to know the
LDOS at all energies, we need this information only at one position. 
A method such as the Kernel Polynomial Method (KPM) is therefore more appropriate and was adopted in this work.
The KPM is described in detail in Weisse et al. \cite{Weisse06}. 

The KPM comprises two main elements. The first element is
a Chebyshev polynomial expansion of the relevant function, in our case the LDOS. The important parameter is the order $N$ of this expansion.
Also, since the Chebyshev polynomials are defined on the interval $[-1,+1]$ it is necessary to re-scale the original Hamiltonian $H$ so that
it's spectrum is contained within this interval, i.e. to work with $\tilde H$ where
\begin{equation}
    \tilde H = \frac{H}{a}\,.
\end{equation}
We set
\begin{equation}
    a = 14.3 > \frac{D + W_{\rm c} \left (E_{\rm F} = 0 \right)}{2}\,,
\end{equation}
where $D=12$ is the bandwidth of the clean system, i.e. the bandwidth of Eq.~(\ref{Hamiltonian}) with disorder parameter $W=0$.

The second element is a convolution of the expansion with a kernel function. This has the effect of smoothing the function that is being expanded.
We use the Jackson kernel. 
The result is that the delta function in the definition Eq. (\ref{LDOS}) of the LDOS is replaced with a function that is approximately Gaussian with
a width  that is approximately equal to $\pi/N$ near the centre of the spectrum and equal to $\pi/N^{3/2}$ near the edge of the spectrum.
Since we set the Fermi energy at the band centre, in what follows the former expression is more relevant.

The integral in Eq.~(\ref{Ftk}) was approximated using Gauss-Chebyshev quadrature. The abscissa are given in Eq. (82),  and the approximation of the integral in Eq. (90)
of Ref.\cite{Weisse06}.
The number of abscissa was set to twice the number of moments.
The number of moments required depends on the Kondo temperature $T_{\rm K}$, which is not known in advance.
Therefore, we adopted an iterative procedure.
For a given sample, we first performed the calculation with a relatively small number of moments. The number of
moments was then multiplied by eight, the calculation repeated, and the Kondo temperature
found compared with that found at the previous iteration.
This was continued until either the Kondo temperature converged to a non-zero
value or a maximum number of moments was reached. 
The smallest Kondo temperature that can be resolved 
reliably is of the order of the resolution $a\pi/N$ of the KPM .
Since the exponent of the power law is expected to change for Kondo temperatures below the level spacing,
we set the maximum of the number of moments such that the resolution of the KPM 
is of the order of the level spacing, giving 
\begin{equation}
    N \approx a \pi L^3/D \;.
\end{equation}

The Kondo temperature varies over many orders of magnitude,
so we solved Eq.~(\ref{NagaokaSuhl}) 
by first changing the variable to $z=\ln T_{\rm K}$
and then applying Brent's method\cite{NumericalRecipes} to the equation
\begin{equation}
    1=\tilde{F} \left(z\right)\,,
\end{equation}
where
\begin{equation}
    \tilde{F} \left( z \right) = F\left( e^z \right)\,.
\end{equation}
The tolerance in Brent's method was set to $10^{-7}$.
When no non-zero solution could be found for the maximum number
of moments in the KPM, we assumed that this indicated a zero
Kondo temperature, i.e. a free magnetic moment.
This overestimates the number of free moments since we cannot then
distinguish a finite Kondo temperature below the level spacing from zero.

The necessary computations were performed on System B of the Institute of Solid State Physics at the University of Tokyo.
Samples were simulated in parallel using MPI on 288 nodes.
The total amount of processor time used was approximately 640 hours.

\section{Results}

\begin{table}[t]
\caption{The exchange coupling $J$, the system size $L$,
the number $N_{\rm s}$ of samples simulated, 
the Kondo temperature of the clean system $T^{(0)}_K$,
the ratio of $T^{(0)}_K$ to the level spacing, and
the maximum number $N$ of moments used in the KPM.}
\label{tab:parameters}  
\begin{tabular}{llllll}
\hline\noalign{\smallskip}
$J$ & $L$ & $N_{\rm s}$ & $T_{\rm K}^{(0)}$ & $T_{\rm K}^{(0)} / \Delta$ & $N$ \\
\noalign{\smallskip}\hline\noalign{\smallskip}
4 & 64 & 100,800 & 0.6473579 & 14,000 & 1,048,576 \\
4 & 96 & 65,664  & 0.6473579 & 47,000 & 4,194,304 \\
6 & 64 & 100,800 & 1.2224582 & 26,000 & 1,048,576 \\
\noalign{\smallskip}\hline
\end{tabular}
\end{table}
  
\begin{figure}
\includegraphics[width=0.45\textwidth]{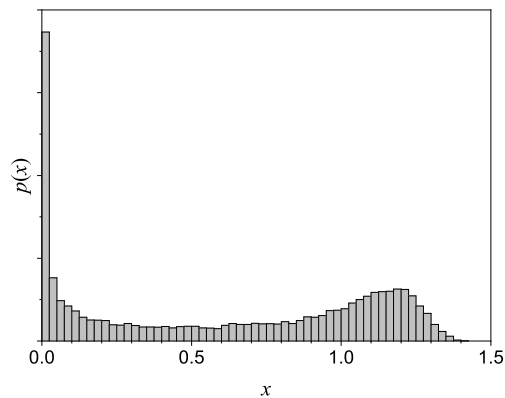}
\caption{The distribution of $x=T_{\rm K} / T_{\rm K}^{(0)}$ 
found for exchange coupling $J=4$ and system size $L=96$.}
\label{fig1} 
\end{figure}

\begin{figure}
\includegraphics[width=0.45\textwidth]{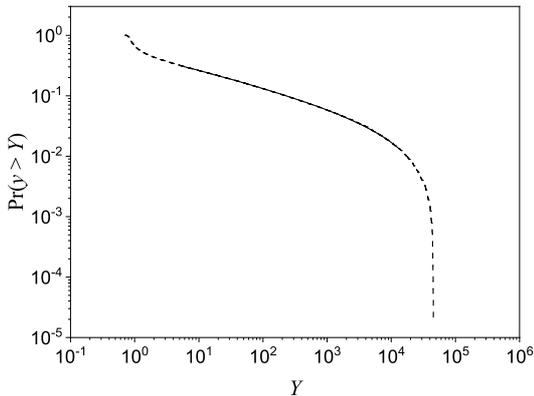}
\caption{The probability ${\rm Pr}(y>Y)$ that the quantity 
$y=T_{\rm K}^{(0)} / T_{\rm K}$ exceeds the value $Y$. The dashed lined shows this probability for the simulation data with $J=4$ and $L=64$.
The solid line is the same quantity for the power law fit.}
\label{fig2}
\end{figure}

\begin{figure}
\includegraphics[width=0.45\textwidth]{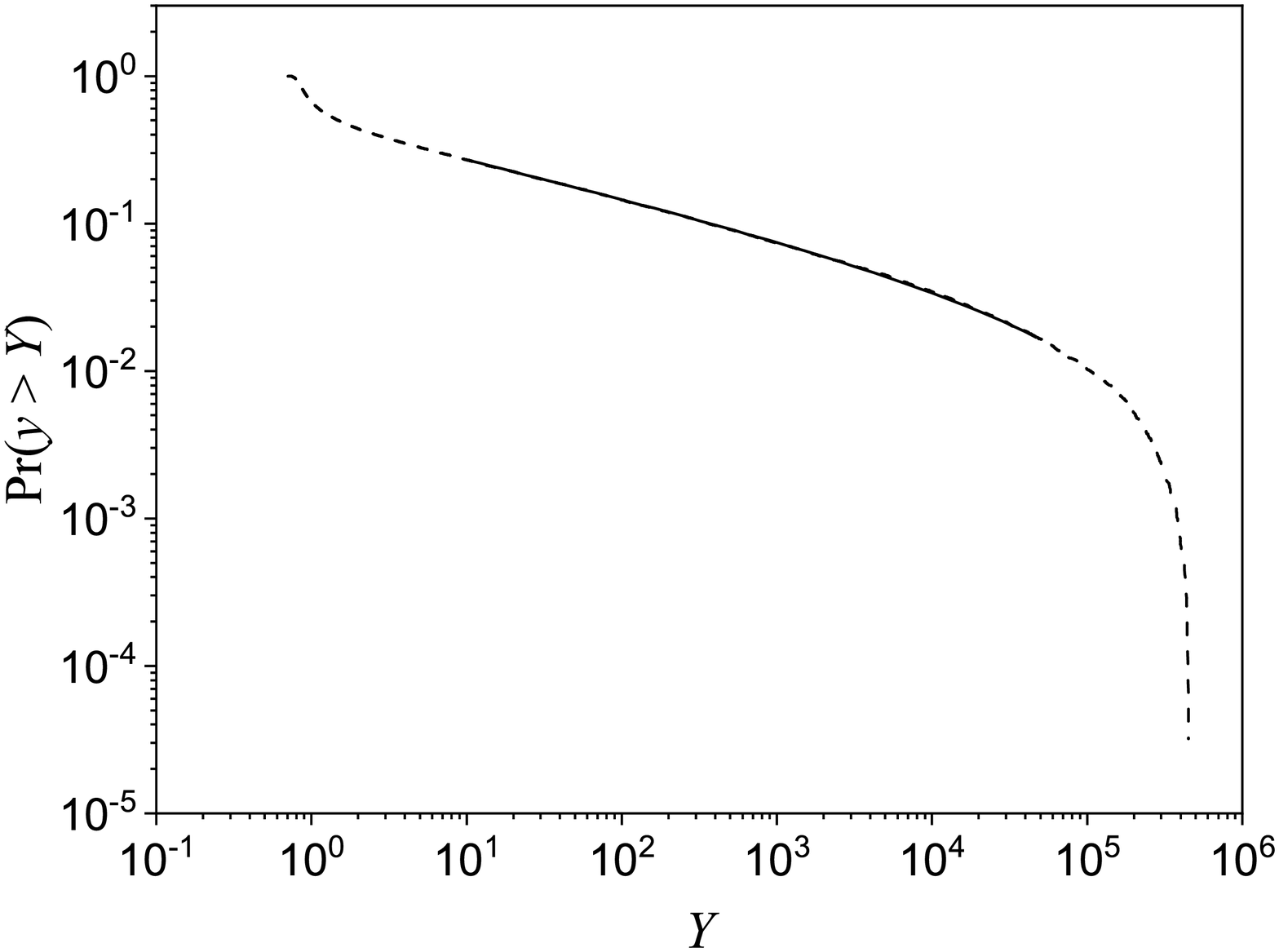}
\caption{The same as Fig. \ref{fig2} for the simulation data with $J=4$ and $L=96$.}
\label{fig3} 
\end{figure}

\begin{figure}
\includegraphics[width=0.45\textwidth]{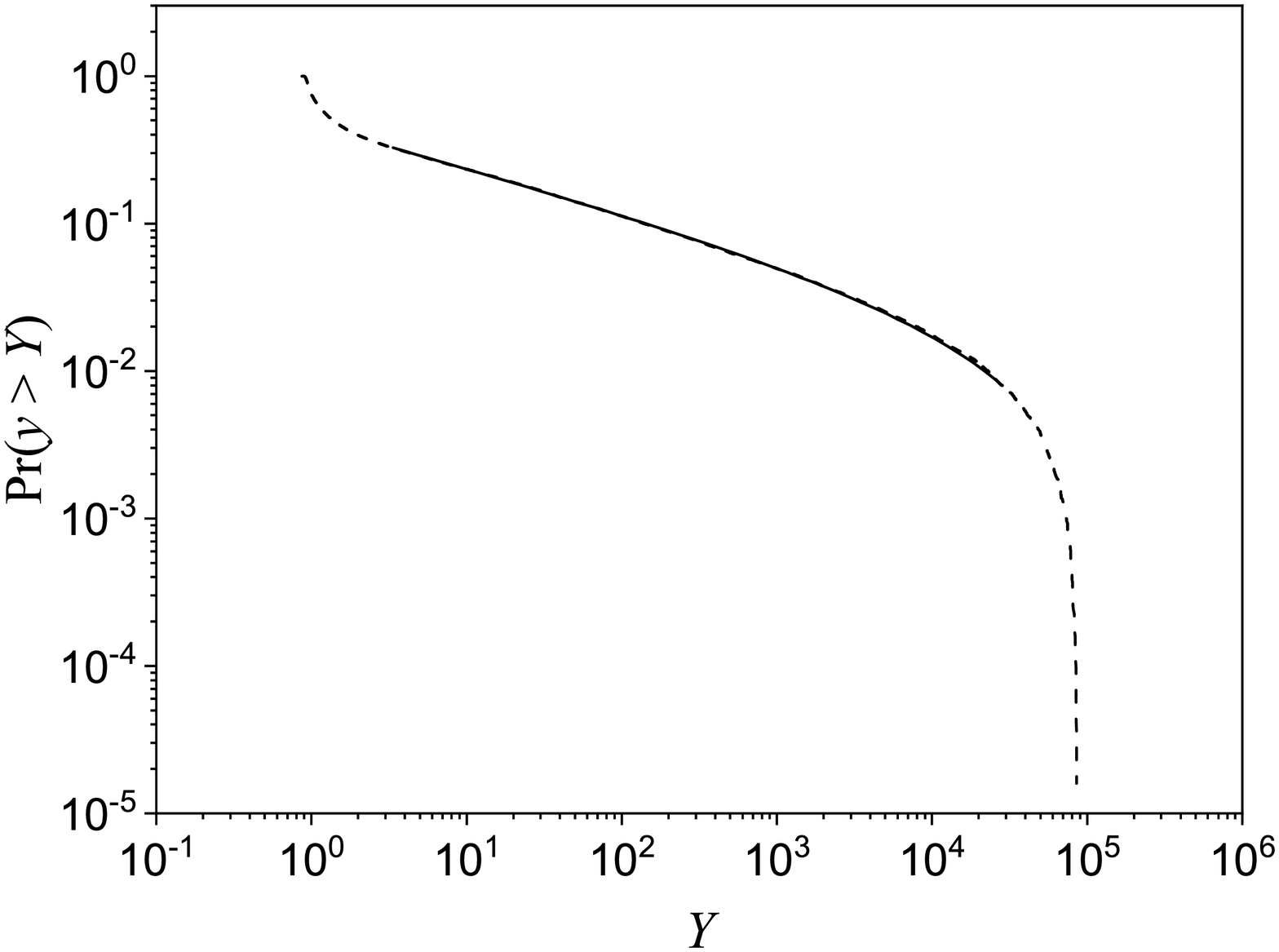}
\caption{The same as Fig. \ref{fig2} for the simulation data with $J=6$ and $L=64$.}
\label{fig4} 
\end{figure}

\begin{figure}
\includegraphics[width=0.45\textwidth]{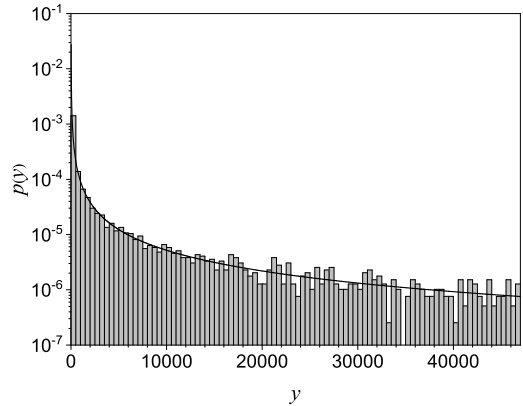}
\caption{The probability density function of the quantity
$y=T_{\rm K}^{(0)} / T_{\rm K}$. 
The histogram is the simulation data for $J=4$ and $L=96$ in the interval 
$(y_{\rm min}, y_{\rm max})$. The solid line is Eq.~(\ref{y_powerlaw}).}
\label{fig5} 
\end{figure}

Three simulations were performed with different sets of parameters. 
The parameters are listed in Table \ref{tab:parameters}.
In this table we also list the Kondo temperature of the clean system.
These were found by solving Eq. (\ref{NagaokaSuhl})
for Anderson's model of localisation with the given $J$ and $L$ with
the disorder $W=0$.

In Figure \ref{fig1} we plot the probability density of the ratio $x$ defined
in Eq. (\ref{x_def}) found for $J=4$  and $L=96$.
There is a peak in the distribution slightly above $x=1$, 
which corresponds to the Kondo temperature of the clean system.
There is also a long tail toward low temperatures.

To analyse the form of the distribution at low temperatures, we found it convenient to
transform to the reciprocal variable
\begin{equation}\label{y_def}
    y = \frac{1}{x} = \frac{T^{(0)}_K}{T_{\rm K}}\,.
\end{equation}
Note that the tail of the distribution is at large $y$.
After transforming to this reciprocal variable, the probability distribution
becomes
\begin{equation}\label{y_powerlaw}
P(y) = p(y)dy = C y^{-\beta} dy\,,
\end{equation}
with the universal exponent
\begin{equation}\label{beta}
    \beta = 1 + \frac{\eta}{2d} \;.
\end{equation}
We expect this to hold for a certain range of data
\begin{equation}\label{y_range}
   y_{\rm min} < y < y_{\rm max} \,.
\end{equation}
Once this range is specified the normalisation constant $C$ is determined
\begin{equation}
   C = \frac{{1 - \beta }}{{y_{\max }^{1 - \beta } - y_{\min }^{1 - \beta }}}\,.
\end{equation}
We determined the value of the lower limit of this range during the fitting of the 
simulation data (see below).
Following Eq. (\ref{temperature_range}), we set the value of the upper limit to
\begin{equation}\label{ymax_def}
    y_{\rm max} = \frac{T^{(0)}_K}{\Delta}\,.
\end{equation}
The values of this upper limit for our simulations are listed in 
Table \ref{tab:parameters}.

To determine if the numerical data are consistent with a power law 
and to estimate the exponent
of the power law,  we followed closely the procedure described by Clauset et al.\cite{Clauset09}.
Those authors  considered the case where the distribution is power law above a 
certain minimum value $y_{\min}$.
Since, for our simulations, there is also an upper limit, we modified their 
procedure to take this into account.
These modifications are described where necessary below. 
For full details of the procedure we refer the reader to Clauset et al.

The procedure has three steps.
The first step is to estimate the exponent $\beta$ and the lower cutoff $y_{\rm min}$.
The maximum likelihood estimate for the exponent $\beta$, i.e. the value of $\beta$
that maximises the probability of the observed data, is the solution of
\begin{equation}
 \frac{1}{{\beta  - 1}} + \frac{{y_{\max }^{1 - \beta }\ln {y_{\max }} - y_{\min }^{1 - \beta }\ln {y_{\min }}}}{{y_{\max }^{1 - \beta } - y_{\min }^{1 - \beta }}} = \frac{1}{n_{\rm t}}\sum\limits_{i = 1}^{n_{\rm t}} {\ln {y_i}}  \,.
\end{equation}
Here, $n_{\rm t}$ is the number of data in the tail of the distribution, 
i.e.  the number of data $y_i$ satisfying Eq.~(\ref{y_range}).
The value of $y_{\min}$ was then varied so as to minimise the 
Kolmogorov-Smirnov statistic,
which measures the discrepancy between cumulative distribution function of
the observed data and the supposed power law distribution.
When determining $y_{\min}$ Clauset et al. perform an exhaustive search over
all the numerical data.
We found that was too time consuming.
Instead, we searched over a set of 100 logarithmically spaced points in the
range $y\in[1,100]$.

The second step is the determination of the goodness of fit.
We did this by generating an ensemble of 10,000 
synthetic data sets as follows.
For each data set we generate $N_{\rm s} - N_{\rm FM}$ 
random numbers between zero and one.
Where these numbers were less than
$n_{\rm t}/(N_{\rm s} - N_{\rm FM})$,
we generated a random number distributed according to Eqs. (\ref{y_powerlaw}) and (\ref{y_range}). 
Otherwise we sampled data with replacement from the simulation data that satisfy
$y\leq y_{\min}$ or $y\geq y_{\max}$.
Each synthetic data set was then subjected to the same fitting procedure as the simulation data.
In this way a distribution for the Kolmogorov-Smirnov statistic was arrived at and 
the goodness of fit determined by comparing the value obtained for the 
fit of the simulation data to this distribution.

The third step was the estimation of the precision of the estimate
of the exponent $\beta$. 
This was done by generating an ensemble of 10,000 synthetic data sets by sampling
the original data with replacement, i.e. the bootstrap method.
Each synthetic data set was then subjected to the same fitting procedure as the simulation data.
In this way a distribution for the exponent $\beta$ was
arrived at and the standard error estimated in the usual way.

The results of this three step procedure for each simulation are given in
Table \ref{tab:results}.
For $J=4$ and $L=64$ and the fit to a power law was successful and an estimate of the exponent obtained. 
For $J=4$ and $L=96$ and the goodness of fit is somewhat lower but the results of the fit
are consistent with the simulation of the smaller system size.
For $J=6$ the fit was not successful.
For this case we report nominal values of the fitting parameters but it should be borne in
mind that their meaning is questionable since the goodness of fit is too small. 

\begin{table}
\caption{The results of the power law fit to the tails of the distribution. 
The exchange coupling $J$, the system size $L$, the estimate $\beta$ of the exponent 
and its standard deviation $\sigma_{\beta}$ (where available). 
Also the goodness of fit probability (GOF)
and the number of data $n_{\rm t}$ in the tail of the distribution,
i.e., the number of data that satisfy Eq. (\ref{y_range}).
In the last row, the restriction $y_{\min}>100$ was imposed when performing the fit.}
\label{tab:results} 
\begin{tabular}{lllllll}
\hline\noalign{\smallskip}
$J$ & $L$ & $\beta$ & $\sigma_{\beta}$ & GOF & $y_{\rm min}$ & $n_{\rm t}$\\
\noalign{\smallskip}\hline\noalign{\smallskip}
4 & 64 & 1.251 & 0.007 & 0.1    & 5.6   & 13,895 \\ 
4 & 96 & 1.24  & 0.01  & 0.002  & 10.2  & 7,866  \\ 
6 & 64 & 1.29  & -     & 0      & 12.9  & 13,041 \\ 
6 & 64 & 1.25  & -     & 0      & 443   & 3,612 \\ 
\noalign{\smallskip}\hline
\end{tabular}
\end{table}

To give a graphical impression of the fits we plot on logarithmic scales the probability (dashed line)
\begin{equation}
    {\rm Pr}(y>Y) = \int_{Y}^{\infty} p(y) dy \,,
\end{equation}
that $y$ exceeds the value $Y$ in Figs. \ref{fig2}, \ref{fig3}, and \ref{fig4}.
We compare this with the power law fit for the relevant range (solid line).
In all cases, the fit is over 3 orders of magnitude of the abscissa.

In Fig. \ref{fig5} we also plot the distribution of $y$ obtained in one
simulation in a more conventional manner (histogram).
The power law fit obtained as described above is also plotted (solid line).
We emphasise that solid line in the figure is {\it not} obtained by fitting
the histogram.
Clauset et al.\cite{Clauset09} report that the often employed method of fitting 
a histogram to a power law on a logarithmic scale sometimes gives biased results.

In Table \ref{tab:freemoments} we list the number of free moments found in 
each numerical simulation.
We also express this as a fraction.

\begin{table}[t]
\caption{The exchange coupling $J$, the system size $L$,
the number $N_{\rm FM}$ of free moments found and
the corresponding fraction $n_{\rm FM}$ of free moments.}
\label{tab:freemoments}
\begin{tabular}{llll}
\hline\noalign{\smallskip}
$J$ & $L$ & $N_{\rm FM}$  & $n_{\rm FM}$ \\
\noalign{\smallskip}\hline\noalign{\smallskip}
4 & 64 & 53,803 & $53\%$  \\
4 & 96 & 34,366 & $52\%$  \\
6 & 64 & 38,204 & $38\%$  \\
\noalign{\smallskip}\hline
\end{tabular}
\end{table}

\section{Discussion}

The derivation of the analytic approximation for the distribution of
Kondo temperatures involves two important approximations.
One is that only pairwise correlations of the wavefunction intensities are included.
Another is the parabolic approximation for the multifractal spectrum.
The main purpose of the numerical simulations and analysis reported here
is to check that the prediction of a power law tail for the distribution of
Kondo temperatures with a universal exponent
holds regardless of these approximations.
Using the value for $\eta$ in Eq. (\ref{eta_parabolic_value}), 
the expected value of the exponent is
\begin{equation}\label{beta_parabolic_value}
    \beta \approx 1.348 \,.
\end{equation}
In our opinion, the successful fitting of the data for $J=4$ to a power law
with an exponent in reasonable, if not perfect, agreement with this
predicted value is evidence that this is the case (see Table \ref{tab:results}).

The failure of the fit to the data for $J=6$ remains to be explained.
However, it may be related to the fact that the temperature range over which
the power law Eq. (\ref{x_powerlaw}) is a good approximation to Eq. (\ref{x_pdf})
depends on $J$ and is more restricted for $J=6$ than $J=4$.
We attempted to check this by fitting the data for $J=6$ subject to the restriction
that $y_{\min}>100$. 
(In this case the search for $y_{\min}$ was performed for 100 logarithmically spaced points in $[100,1000]$.)
The results are shown in the last row of Table \ref{tab:results}.
While the goodness of fit was still not acceptable we did notice that the nominal value of
the exponent is now in agreement with that found for $J=4$.

The values for the exponent found in the numerical simulations (see Table \ref{tab:results}) are slightly smaller than that given in Eq. (\ref{beta_parabolic_value}).
While the derivation of the analytic approximation for the distribution of Kondo
temperatures that leads to Eq. (\ref{beta}) involves the parabolic approximation, we may
speculate that this relation might still hold with the exact value of $\eta$.
Using the numerical value of $D_2$ derived from Ref. \cite{Rodriguez11} we find
\begin{equation}
    \eta \approx 1.763 \,,
\end{equation}
and 
\begin{equation}
    \beta \approx 1.294 \,.
\end{equation}
This is in better but still not perfect agreement with the value found in the 
numerical simulations.

For the fraction of free moments, the values obtained with Eq. (\ref{nfm}) are of the order
of several percent.
Even allowing for the fact that, because of the finite level spacing, our simulations overestimate
the fraction of free moments, 
this is much less than found in the numerical simulations (see Table \ref{tab:freemoments}).
While the analytic estimation of the fraction of free moments is
much more sensitive to the parabolic approximation for the multifractal spectrum
than the estimation of the exponent $\beta$ it seems difficult to fully account for the discrepancy on this basis and this remains a puzzle.

\begin{acknowledgements}
The authors thank the Supercomputer Center, 
Institute for Solid State Physics, University of Tokyo for the use of System B. 
This work was supported by JSPS 
KAKENHI grants 19H00658 (K.S. and T.O.) and 26400393 (K.S.),
and DFG grant KE 807/22-1 (S.K.).

\end{acknowledgements}

\bibliographystyle{spphys}        
\bibliography{references}   

\end{document}